\documentclass[pra,aps,superscriptaddress,showpacs,reprint,longbibliography]{revtex4-1}

\usepackage{etoolbox}
\usepackage{gensymb}
\usepackage{hyperref}
\usepackage{amsmath}
\usepackage{amssymb}
\usepackage{dsfont}
\usepackage[T1]{fontenc}
\usepackage{ae}
\usepackage{color}
\usepackage{graphicx} 
\usepackage{units}
\usepackage{placeins}

\begin{document}

\title{Applying Electric and Magnetic Field Bias in a 3D Superconducting Waveguide Cavity with High Quality Factor}

\author{M. Stammeier} \email{mathiass@phys.ethz.ch} \affiliation{Department of Physics, ETH Z\"urich, CH-8093 Z\"urich, Switzerland}

\author{S. Garcia} \affiliation{Department of Physics, ETH Z\"urich, CH-8093 Z\"urich, Switzerland}

\author{A. Wallraff} \affiliation{Department of Physics, ETH Z\"urich, CH-8093 Z\"urich, Switzerland}

\renewcommand{\i}{{\mathrm i}} \def\1{\mathchoice{\rm 1\mskip-4.2mu l}{\rm 1\mskip-4.2mu l}{\rm
1\mskip-4.6mu l}{\rm 1\mskip-5.2mu l}} \newcommand{\ket}[1]{|#1\rangle} \newcommand{\bra}[1]{\langle
#1|} \newcommand{\braket}[2]{\langle #1|#2\rangle} \newcommand{\ketbra}[2]{|#1\rangle\langle#2|}
\newcommand{\opelem}[3]{\langle #1|#2|#3\rangle} \newcommand{\projection}[1]{|#1\rangle\langle#1|}
\newcommand{\scalar}[1]{\langle #1|#1\rangle} \newcommand{\op}[1]{\hat{#1}}
\newcommand{\vect}[1]{\boldsymbol{#1}} \newcommand{\id}{\text{id}}

\raggedbottom

\begin{abstract}
Three-dimensional microwave waveguide cavities are essential tools for many cavity quantum electrodynamics experiments. However, the need to control quantum emitters with dc magnetic fields inside the cavity often limits such experiments to normal-conducting cavities with relatively low quality factors of about $10^4$. Similarly, controlling quantum emitters with dc electric fields in normal- and superconducting waveguide cavities has so far been difficult, because the insertion of dc electrodes has strongly limited the quality factor. Here, we present a method to apply dc electric fields within a superconducting waveguide cavity, which is based on the insertion of dc electrodes at the nodes of the microwave electric field. Moreover, we present a method to apply dc magnetic fields within the same cavity by trapping the magnetic flux in holes positioned in facing walls of the cavity. We demonstrate that the $\text{TE}_{301}$ mode of such a superconducting, rectangular cavity made from niobium maintains a high internal quality factor of $Q_{\text{int}} \sim 1.7 \cdot 10^6$ at the few photon level and a base temperature of $3~\text{K}$. A cloud of Rydberg atoms coupled to the microwave electric field of the cavity is used to probe the applied dc electric and magnetic fields via the quadratic Stark effect and the Zeeman effect, respectively. 
\end{abstract} 	
\maketitle 
  
\section{Introduction}
Three-dimensional (3D) microwave resonators play an important role in cavity quantum electrodynamics (QED)~\cite{Haroche2006} and for applications in quantum information processing~\cite{Chuang2010}. Superconducting Fabry-Perot resonators with photon lifetimes of 100 milliseconds have been extensively used in cavity QED experiments with Rydberg atoms~\cite{Haroche2013}. While their open structure (gap between the two mirrors) allows the application of dc electric and magnetic fields within the cavity, the design and fabrication is considerably more demanding than, e.g., for waveguide resonators of rectangular and cylindrical shape. Such (closed) waveguide cavities have in recent years attracted considerable interest in the context of superconducting quantum circuits  because of increased coherence times~\cite{Paik2011,Rigetti2012} and long single photon lifetimes on the order of milliseconds~\cite{Reagor2013}, which has enabled key experiments in circuit QED~\cite{Sun2014,Kirchmair2013,Gasparinetti2016} and led to new directions in quantum information processing~\cite{Flurin2015,Heeres2015,Reagor2016}. 
More recently, the homogeneous spatial distribution of microwave fields inside waveguide 3D resonators was exploited to demonstrate strong coupling between photons and spin ensembles in nitrogen vacancy centers~\cite{Angerer2016}, in a YIG sphere~\cite{Zhang2014} and in rare-earth-doped crystals~\cite{Probst2014}, which enabled the realization of a hybrid quantum system between superconducting qubits and magnons~\cite{Tabuchi2015}. 

 Many quantum emitters that can be coupled to waveguide cavities share a common feature: their transition frequencies can be controlled by either dc magnetic fields (superconducting qubits and spin ensembles), dc electric fields (nanomechanical resonators~\cite{Singh2014} for optomechanics~\cite{Yuan2015}) or both (nitrogen vacancy centers~\cite{Dolde2011} and Rydberg atoms~\cite{Hare1988}). 
 However, the application of dc magnetic fields in superconducting waveguide cavities is challenging, since magnetic field lines can not penetrate the walls of a superconducting cavity due to the Meissner effect. One solution to this problem is a hybrid cavity~\cite{Reshitnyk2016}, where a copper insert facilitates the penetration of magnetic fields, but also reduces the cavity's quality factor. The incorporation of dc electric field bias into a superconducting 3D cavity is considerably more challenging. Indeed, each resonator mode has a characteristic distribution of surface currents on the inner walls of the cavity and a change of this distribution, e.g.~by introducing a hole for an electrode, can alter the resonator mode itself. Furthermore, electrodes inserted into a cavity can cause substantial radiation losses, because they act as antennas that couple to the microwave field inside the cavity. Initial attempts to position multiple on-chip dc bias lines into a rectangular 3D cavity~\cite{Kong2015} have shown a large reduction of the cavity's quality factor down to about $10^3$. Cohen et al.~\cite{Cohen2017} have recently presented a method that circumvents the need for intra-cavity electrodes and allows the application of a single dc voltage bias in a superconducting 3D cavity. In their study they used two isolated halves of a split, rectangular cavity as electrodes and achieved a high internal quality factor of $9 \cdot 10^5$ at the single photon level.

Here, we present a method for incorporating dc bias voltages into a 3D cavity, which is based on the insertion of dc electrodes at the nodes of the microwave electric field of a superconducting cavity. We further devise a way of applying static magnetic fields using two holes in opposite walls of the cavity, where the magnetic flux can be trapped when  the cavity is cooled down in an external magnetic field. The combination of both methods enables the study of quantum emitters  coupled to the microwave electric field of a superconducting cavity in dc electric and magnetic fields. We demonstrate that the $\text{TE}_{301}$ mode of a superconducting, rectangular cavity with two inserted dc electrodes and two holes for magnetic field application maintains a high internal quality factor of $\simeq 1.7 \cdot 10^6$ at the few photon level, as measured at $T=3~\text{K}$. We prove the robustness of the methods against imperfections, e.g.~limited precision in positioning of the dc electrodes, with the insertion of a high-purity sapphire cylinder, which affects the electric and magnetic field distribution of the cavity mode used, while the quality factor is unaltered.

\begin{figure}[b] \centering \includegraphics[width=85mm]{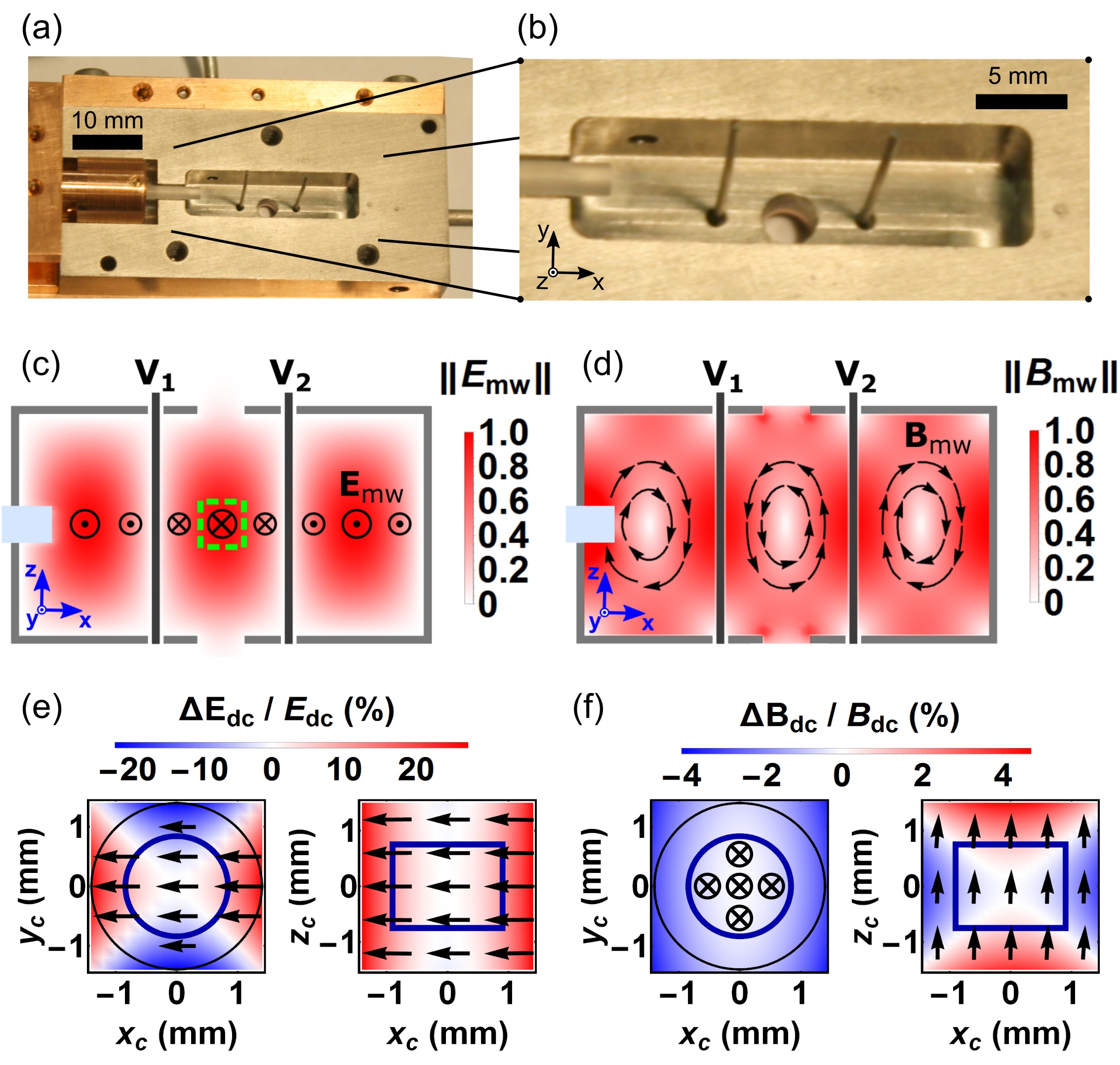} 	
	\caption{(a) Photograph of one half of the rectangular niobium 3D cavity in its copper mounting bracket also used for thermalization. Variable insertion depth of the low-microwave-loss sapphire cylinder on the left into the cavity permits the tuning of its  resonance frequencies. (b) Detail from (a): Inside the cavity two niobium dc electrodes are mounted at the nodes of the electric field of the $\text{TE}_{301}$ mode. A 3-mm-diameter access hole (see text) is centered in the cavity wall. The 1-mm feedthrough in the upper cavity wall hosts one of two microwave evanescent couplers.  
	(c,d) The simulated $\text{TE}_{301}$ mode's electric (c) and magnetic (d) field amplitude distribution plotted in the plane of the cavity electrodes. Black arrows indicate the orientation of the microwave field vectors $\boldsymbol{E}_{\text{mw}}$ and $\boldsymbol{B}_{\text{mw}}$.
	(e,f) Simulation of the dc electric (e) and magnetic (f) fields, $\boldsymbol{E}_{\text{dc}}$ and  $\boldsymbol{B}_{\text{dc}}$, in a small region (green dashed box in (c)) around the cavity center (coordinates $x_{\text{c}}$, $y_{\text{c}}$ and  $z_{\text{c}}$ relative to the latter). The simulated field vectors (black) are plotted on top of the relative deviations, $\Delta E_{\text{dc}} / E_{\text{dc}}$ and $\Delta B_{\text{dc}} / B_{\text{dc}}$, of the field magnitudes from the values in the center, $E_{\text{dc}}$  and $B_{\text{dc}}$ (see text for details). Black circles represent access holes and blue lines indicate the region, where field distributions are probed by Rydberg atoms (see text).}
	\label{fig:Fig1}
\end{figure}

\section{Cavity design and field simulations}
In Fig.~\ref{fig:Fig1}(a,b), we show one of the two symmetric, rectangular cavity halves, which are milled out of high-purity ($RRR = 300$) niobium, aligned with stainless steel posts mounted in the cavity walls and fixed to each other by a set of stainless steel screws. The cavity dimensions ($L_x \times L_y \times L_z \simeq 25.6~\text{mm} \times 7~\text{mm} \times 14~\text{mm}$) determine the resonance spectrum with the lowest frequency transverse electric modes $\text{TE}_{101}$, $\text{TE}_{201}$ and $\text{TE}_{301}$. Two cylindrical dc electrodes with a diameter of 0.5~mm are inserted at the electric field nodes of the $\text{TE}_{301}$ mode and allow the application of dc electric fields inside the cavity. Each dc electrode is formed by a microwave coaxial cable, the stripped center conductor of which enters the cavity, while an additional length  of coaxial cable (about $30~\text{mm}$) is clamped with a copper plate outside the cavity for thermalization.   
Two 3-mm-diameter access holes, one in each cavity half, are used to trap dc magnetic fields and allow Rydberg atoms to enter and exit the cavity. Additionally, the cavity features a 2.3-mm-diameter hole on the left side, through which we insert a 1.9-mm-diameter cylinder of either low-microwave-loss sapphire (shown) or superconducting niobium, with a variable depth set by a translation stage. Two microwave evanescent couplers (retracted in the holes used as a feedthrough, see Fig.~\ref{fig:Fig1}(b)), symmetrically coupled to the cavity, are used to inject microwave radiation into and extract it from the cavity in transmission measurements.

 At the electric field nodes the $\text{TE}_{301}$ mode is compliant with conducting boundary conditions imposed by the dc electrodes mounted in the cavity. As a consequence we expect a minimal perturbation of the mode's field distribution. To confirm this expectation, we perform finite-element simulations~\cite{HFSS2014} of the cavity fields with a sapphire rod inserted up to a depth of $\delta x_{\text{S}}=2~\text{mm}$ (distance between the tip of the sapphire and wall) into the cavity. The simulated electric and magnetic field amplitude distributions plotted in the plane of the electrodes (see Figure~\ref{fig:Fig1}(c,d)) closely resemble the ideal distributions, e.g.~$E_{\text{mw}}=|\sin(3\pi x /L_x ) \sin (\pi z /L_z )|$ for the electric field. A quantitative comparison between the simulated and the ideal field distributions shows that the introduction of the sapphire cylinder, access holes and, most importantly, dc electrodes cause less than $\simeq 8~\%$  relative deviations in the electric and magnetic field amplitudes at all positions separated by more than $2~\text{mm}$ from the listed perturbing objects.

  We then study spatial distributions of applied dc electric and magnetic fields by finite element simulations~\cite{Maxwell2014} in a small region around the cavity center (see  Fig.~\ref{fig:Fig1}(e,f)). The electric field is simulated with potential differences $V_1=-1~\text{V}$ and $V_2=1~\text{V}$ of the first and second cavity electrode to the cavity, respectively. It is oriented in x-direction and reasonably homogeneous with less than $10\,\% $ averaged modulus of relative deviation of the field magnitude from the value in the cavity center $E_{\text{dc}}= 0.95~\text{V}/\text{cm}$. Realistic simulations of magnetic fields in the presence of type 2 superconductors, like niobium, are challenging due to vortex trapping~\cite{Aull2012}, which leads to non-zero magnetic field in the bulk material. Here, we consider the worst case scenario for the homogeneity of the magnetic field by choosing a perfectly diamagnetic material for the cavity, which models a type 1 superconductor. 
  Indeed, in this case, all magnetic field lines enter and leave the cavity through the access holes; their dilution inside the cavity leads to an inhomogeneous magnetic field.
  In an external magnetic field of $10~\text{G}$ pointing in z-direction, the simulation still predicts an almost homogeneous magnetic field at the cavity center with less than $\pm 4\,\% $ relative deviation of the field magnitude from the center value $B_{\text{dc}}= 4.50~\text{G}$. 

 %at position $\boldsymbol{x}_{0}$ using relative coordinates,  $\boldsymbol{x}_{\text{c}}= \boldsymbol{x} - \boldsymbol{x}_{0}$
%Data and fits for the cavity with dc electrodes are frequency shifted by $12~\text{MHz}$ ($\text{TE}_{101}$), $32~\text{MHz}$ ($\text{TE}_{201}$) and $-41~\text{MHz}$ ($\text{TE}_{301}$) with respect to the resonance frequencies of the empty cavity indicated at the top axis.

\section{Losses induced by dc electrodes}

\begin{figure}[t] \centering \includegraphics[width=85mm]{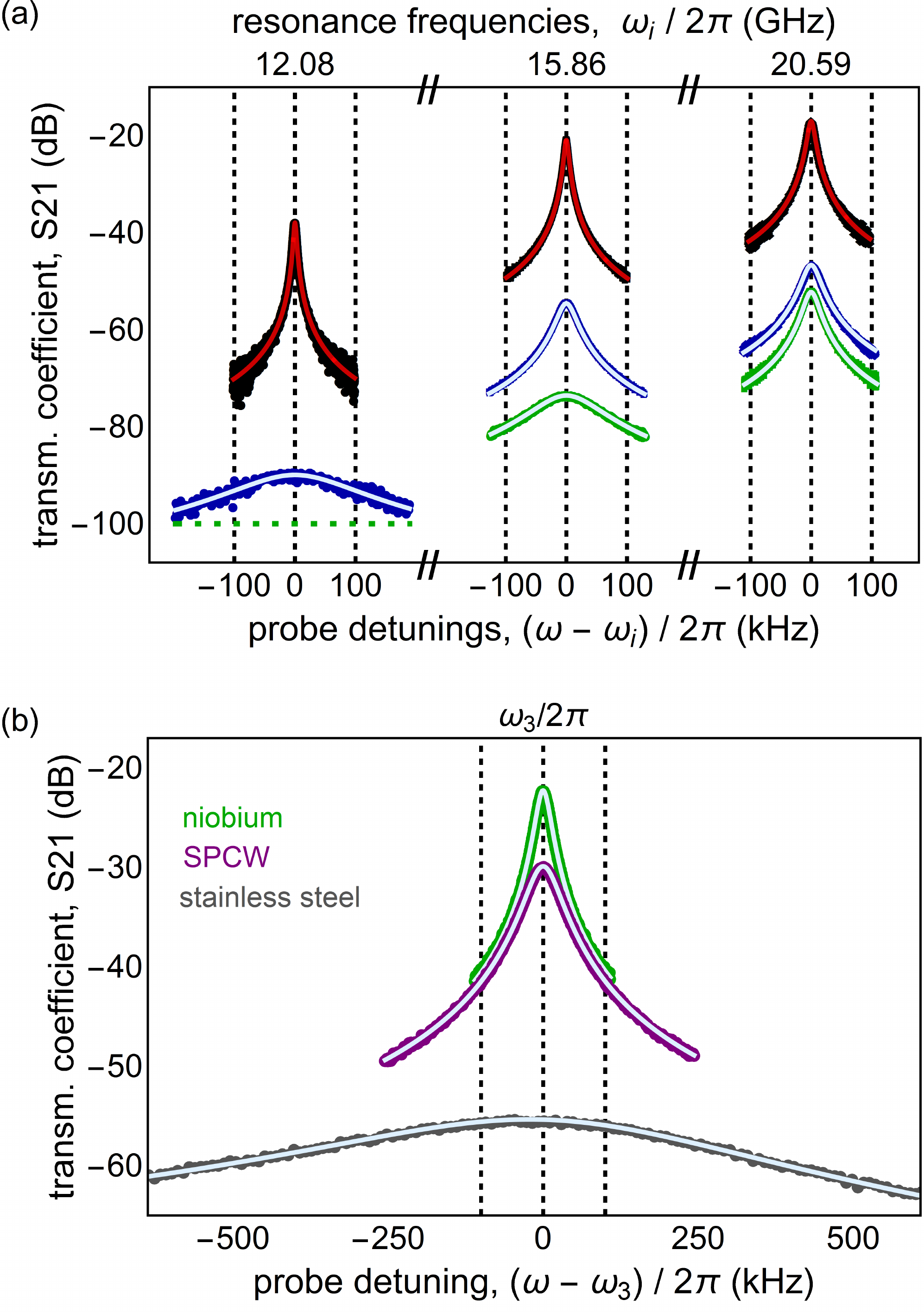} 	
	\caption{(a) Measured S21 transmission spectra of the undercoupled 3D cavity at a temperature of $4.5~\text{K}$ without (black) and with the superconducting dc electrodes installed (green, offset by -30 dB). An additional dataset (blue, offset by -20 dB) at a temperature of $5.2~\text{K}$ was taken after unmounting and remounting the dc electrodes. Lorentzian fits of the transmission spectra for the cavity without (red) and with the dc electrodes mounted (white). Resonance frequencies of the empty cavity are displayed on the top axis. The green dashed line indicates that the fundamental mode could not be measured here.
	(b) Transmission spectra of the $\text{TE}_{301}$ mode at $4.5~\text{K}$ for the undercoupled 3D cavity with superconducting niobium (green), silver-plated copper-weld (SPCW, purple) and stainless steel (gray) dc electrodes. Lorentzian fits are plotted in white. }
	\label{fig:Fig2}
\end{figure}

We experimentally investigate the effect of the superconducting dc electrodes on the resonator modes. We use a vector network analyzer to perform transmission measurements on the cavity without frequency tuning cylinder in an evacuated dipstick at temperatures between $4.5~\text{K}$ and $5.2~\text{K}$.
We subtract the frequency-dependent losses of the microwave cables from the transmission spectra of the $\text{TE}_{101}$, $\text{TE}_{201}$ and $\text{TE}_{301}$ modes of the empty cavity with the access holes and fit the resulting spectra (see black points in Fig.~\ref{fig:Fig2}(a)) with Lorentzians. We thereby obtain the resonance frequencies $\nu_1 =12.08~\text{GHz}$, $\nu_2 =15.86~\text{GHz}$ and  $\nu_3 =20.59~\text{GHz}$ and the cavity linewidths $\kappa_1/(2\pi) = 5.1(1)~\text{kHz}$, $\kappa_2/(2\pi) = 7.6(1)~\text{kHz}$  and $\kappa_3/(2\pi) = 11.9(1)~\text{kHz}$ for the $\text{TE}_{101}$, $\text{TE}_{201}$ and $\text{TE}_{301}$ mode, respectively. The extracted linewidths correspond to high quality factors between $Q_3\sim 1.7 \cdot 10^6$ and $Q_1\sim 2.4 \cdot 10^6$, which are almost identical with the internal quality factors ($Q_{3\text{,int}}\sim 1.9 \cdot 10^6$ and $Q_{1\text{,int}}\simeq 2.4 \cdot 10^6$) determined from the insertion losses. We analyze the origin of the loss rates by performing finite-element simulations, which indicate that microwave field leakage through the access holes only causes negligible loss rates of less than $0.3~\text{kHz}$ for the three considered modes. This suggests that the measured quality factors are limited by losses in the cavity walls, e.g.~residual resistivity and surface defects.

We investigated transmission spectra of the cavity with two superconducting dc electrodes, each of which is terminated with a 50-$\Omega$ impedance to avoid unwanted reflections from the end of the cable. 
%: $T = 4.5~\text{K}$, offset by $-30~\text{dB}$ and blue data points: $T = 5.2~\text{K}$, offset by $-20~\text{dB}$)
We show two sets of measurements (green and blue data points in Fig.~\ref{fig:Fig2}(a)), in between which we have unmounted and remounted the dc electrodes, with an estimated positioning accuracy of $0.1~\text{mm}$, to test the reproducibility of the results. In both datasets all three modes have negligible shifts in the resonance frequencies (about two per mille) compared to the empty cavity. The cavity linewidths, however, vary substantially between the datasets, except for the $\text{TE}_{301}$ mode, which shows a narrow, temperature-dependent cavity linewidth of $\kappa_{\text{Nb,}3}/(2\pi) = 24.8(1)~\text{kHz}$ at $4.5~\text{K}$ and $\kappa_{\text{Nb,}3}/(2\pi) = 28.2(2)~\text{kHz}$ at $5.2~\text{K}$ in all datasets. To understand the variation in cavity linewidth of the $\text{TE}_{101}$ and $\text{TE}_{201}$ modes we perform simulations, which show that the antenna coupling between the microwave field in the cavity and the dc electrodes critically depends on their relative alignment.

Then, we study the influence of dissipation, which increases the cavity linewidth by an amount that is proportional to the surface of the dc electrodes and the surface resistivity $R_s=\sqrt{\mu_0 \omega / \sigma}$~\cite{Pozar2011}. The latter depends on the microwave frequency $\omega$ and the electrode conductivity $\sigma$, $\mu_0$ is the vacuum permeability. In Fig.~\ref{fig:Fig2}(b) we compare transmission spectra of the $\text{TE}_{301}$ mode for the cavity with 50-$\Omega$-terminated dc electrodes of same dimensions made from niobium (Nb, green points), silver-plated copper weld (SPCW, purple points) and stainless steel (SS, gray points). The extracted cavity linewidths, $\kappa_{\text{Nb,}3}/(2\pi) = 24.8(1)~\text{kHz}$, $\kappa_{\text{SPCW,}3}/(2\pi) = 58.2(2)~\text{kHz}$ and $\kappa_{\text{SS,}3}/(2\pi) = 649(3)~\text{kHz}$ show that the cavity losses depend on the conductivity of the electrode material. 
To better understand the data we perform simulations, in which the dissipation on the electrode surface is the only loss mechanism, and obtain the scaling of the electrode losses $\Delta \kappa_{\sigma\text{,}3}/(2\pi) = 121(2)~\text{kHz} / \sqrt{\sigma / (5.8 \cdot 10^7~\text{S}/\text{m}) }$ with the electrode conductivity for the $\text{TE}_{301}$ mode.
Assuming only dissipative losses we can use the scaling to convert the measured increase in cavity linewidth, e.g.~$\Delta \kappa_{\text{SS,}3}/(2\pi) =(\kappa_{\text{SS,}3} - \kappa_{3})/(2\pi)$, to obtain conductivities for stainless steel $\sigma_{\text{SS}} \simeq 2.1\cdot 10^6~\text{S}/\text{m}$ and silver-plated copper weld $\sigma_{\text{SPCW}} \simeq 4.0\cdot 10^8~\text{S}/\text{m}$ at $T\simeq 5~\text{K}$, which are, respectively, compatible with the literature~\cite{Timmerhaus1977} and a residual resistance ratio of $RRR\simeq 7$.

 % The dissipation relates to eddy currents on the electrode surface, which are induced by the tangential component $H_{\text{t}}$ of the microwave magnetic field (see Fig.~\ref{fig:Fig1}(d)). The dissipated power is then given by 
%\begin{equation}
%\int_{A}R_s \cdot |H_{\text{t}}|^{2} dA , 
%\end{equation}
%where the integral is taken over the electrode surface $A$, and $R_s=\sqrt{\mu_0 \omega / \sigma}$ is the surface resistivity for frequency $\omega$ and electrode conductivity $\sigma$.

\begin{figure}[bt] \centering \includegraphics[width=85mm]{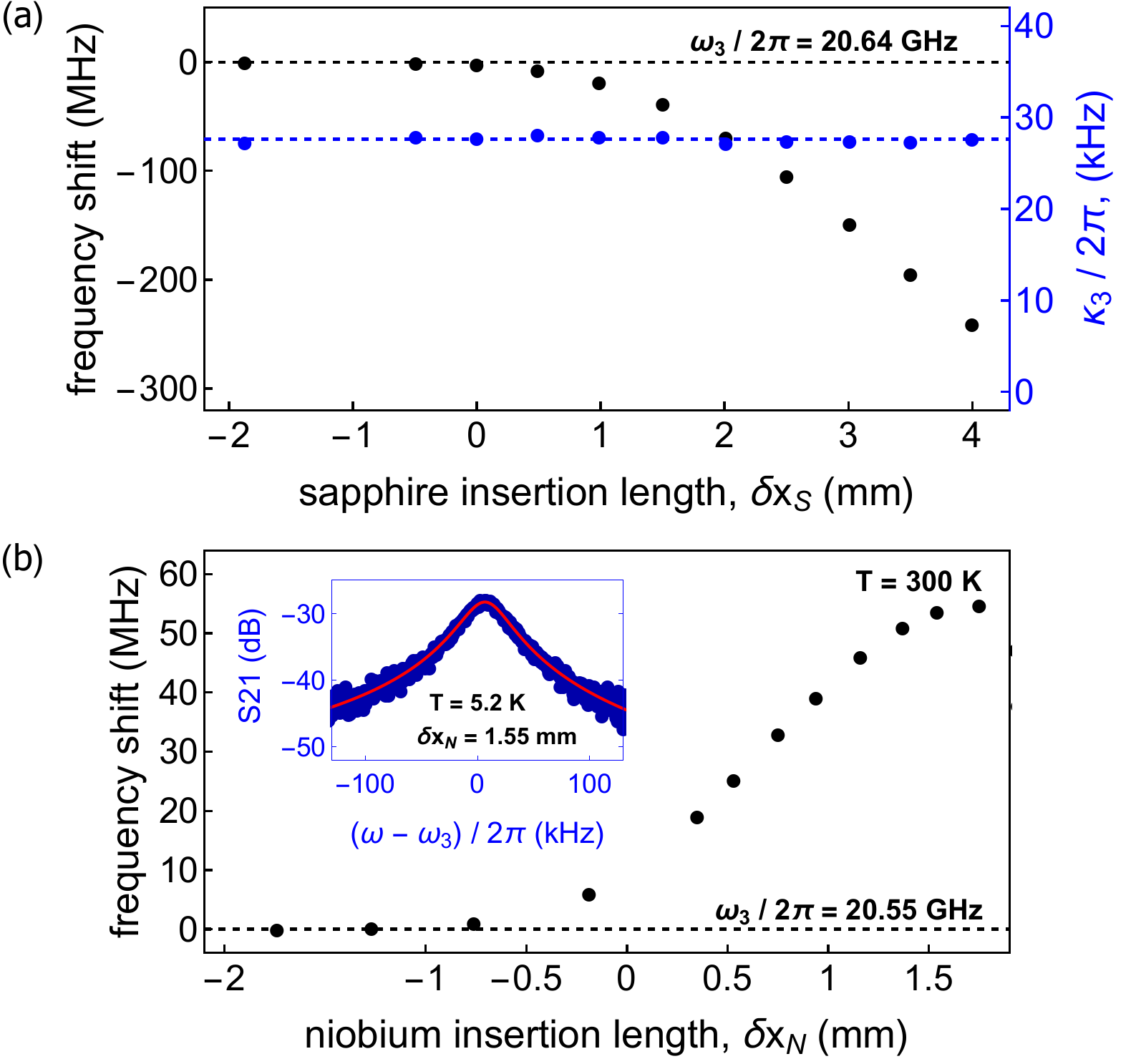} 	
	\caption{Tuning the cavity frequency: (a) Measurements of shift in resonance frequency $\delta \nu_3$ (black data points and left axis) and cavity linewidth $\kappa_3 / 2\pi$ (blue data points and right axis) of the  $\text{TE}_{301}$ mode at $T = 5.2~\text{K}$ versus length $\delta x_{\text{S}}$ of sapphire cylinder inserted into the cavity. (b) Measurement of $\delta \nu_3$  at $T = 300~\text{K}$ versus length $\delta x_{\text{N}}$ of inserted niobium cylinder. The inset shows the power transmission spectrum of the $\text{TE}_{301}$ mode at $T = 5.2~\text{K}$ for a niobium insertion length of $\delta x_{\text{N}}= 1.55 ~\text{mm}$.}
	\label{fig:Fig3}
\end{figure}

Tuning of the cavity frequency is a useful feature~\cite{Carvalho2016,Souris2017} in cavity QED experiments, where the frequency detuning between the emitter and the cavity mode is of fundamental importance. Here, we implement two approaches to tune the resonance frequency of the rectangular waveguide cavity: (i) by inserting a low-microwave-loss sapphire cylinder with dielectric constant $\epsilon_{\text{r}}\simeq 9$ (microwave electric field perpendicular to the c-axis of the crystal), which increases the effective permittivity of the cavity and thereby lowers the resonance frequency; (ii) by inserting a superconducting niobium cylinder, which reduces the cavity volume and thus increases the resonance frequency. Indeed, the cavity frequency decreases monotonically (see Figure~\ref{fig:Fig3}(a)) with increasing insertion length $\delta x_{\text{S}}$ of the sapphire cylinder into the cavity. The sapphire insertion leads to a shift in resonance frequency of up to $\delta \nu_{3\text{,S}}=-230~\text{MHz}$ for $\delta x_{\text{S}}=4.2~\text{mm}$, while the cavity linewidth remains unaffected at $\kappa_{3}=28(1)~\text{kHz}$. We conclude from the measurements that the $\text{TE}_{301}$ mode is robust against losses that may arise from antenna coupling caused by small imperfections in the positioning of the electrodes with respect to the field distribution. In Fig.~\ref{fig:Fig3}(b), we present a room temperature measurement of the shift in resonance frequency $\delta \nu_3$ as a function of the niobium insertion length $\delta x_{\text{N}}$, which shows a positive frequency shift that saturates at  $\delta x_{\text{N}}=1.55~\text{mm}$ with a maximal shift of $\delta \nu_{3\text{,N}}=54~\text{MHz}$. For this insertion length, we have measured the transmission spectrum at $T = 5.2~\text{K}$ (see the inset), which displays a small increase in cavity linewidth, $\kappa_{3}=42(2)~\text{kHz}$ compared to the cavity without inserting the niobium rod. We attribute this increased linewidth to small radiation losses caused by antenna coupling between the niobium rod and the field in the cavity.

\FloatBarrier

%We attribute this increased losses to finite antenna coupling, which is however strongly suppressed by the impedance mismatch between the 3D cavity ($Z_1 \simeq 400~\Omega$) and the coaxial waveguide structure formed by the 1.9-mm-diameter niobium cylinder inside the 2.3-mm-diameter hole ($Z_2 \simeq 10~\Omega$). 

\section{Probing dc electric and magnetic fields with Rydberg atoms}
\begin{figure*}[tb] \centering \includegraphics[width=170mm]{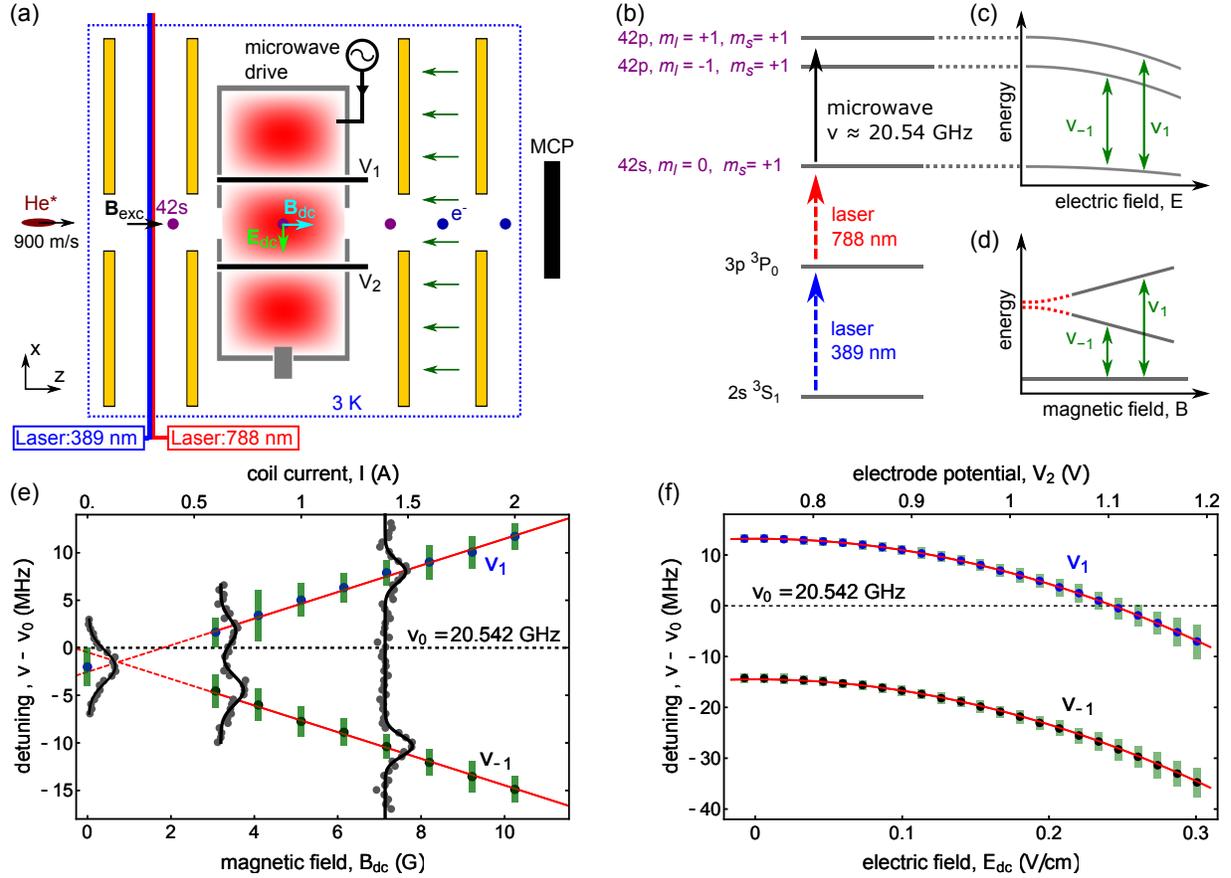} 	
	\caption{(a) Schematic diagram of the experimental setup (not to scale). Within the cryogenic environment at $T=3~\text{K}$ (blue dots) the rectangular 3D cavity (gray) is mounted between two pairs of circular electrodes (yellow). The $\text{TE}_{301}$ mode field amplitude is depicted in red and the embedded electrodes in black. The dc electric and magnetic field vectors, $\boldsymbol{E}_{\text{dc}}$ and $\boldsymbol{B}_{\text{dc}}$, are indicated in green and cyan, respectively. The black arrow illustrates the magnetic field $\boldsymbol{B}_{\text{exc}}$ at the laser excitation. Dark-green arrows represent the pulsed field ionization of the Rydberg atoms, which creates ionization electrons detected on a microchannel plate (MCP). (b) Level diagram for the generation of helium triplet Rydberg states: two-photon excitation to the 42s $m_{l}=0$ state ($S=1$, $m_{s}=1$) by the blue and red laser. A microwave tone, applied to the cavity drives the atoms to the 42p states ($S=1$, $m_{s}=1$) with $m_{l}=1$ and $m_{l}=-1$ at transition frequencies $\nu_{1}$ and $\nu_{-1}$, respectively. (c) Sketch of the quadratic electric field dependence (Stark effect) of the 42s and 42p levels with indication of the transition frequencies  $\nu_{1}$ and $\nu_{-1}$. (d) Sketch of the linear magnetic field dependence of $\nu_{1}$ and $\nu_{-1}$ in the Paschen-Back regime. Red dashed lines indicate the region of small magnetic field (regime of anomalous Zeeman effect).
	(e,f) Measurement of transition frequencies, $\nu_{1}$(blue data points) and $\nu_{-1}$ (black data points), relative to the field-free transition frequency $\nu_{0}$, as a function of the applied magnetic (e) and electric field (f). Full red lines are fits of the Zeeman effect in the Paschen-Back regime to the data ((e), extrapolated with dashed lines) and the quadratic Stark effect (f), which facilitate the calibration between top and bottom axes in both plots. Green bars indicate the full widths at half maximum of the fitted atomic spectral lines, examples of which are displayed in (e) for $B_{\text{dc}}\simeq 0~\text{G}$, $B_{\text{dc}}\simeq 3.1~\text{G}$ and $B_{\text{dc}}\simeq 7.2~\text{G}$ with gray data points and black fits. The measurements in (f) are conducted in a magnetic field of $B_{\text{dc}}\simeq9.8~\text{G}$ and with the other electrode at a constant potential of $V_{1} = 0.6 \text{V}$.  }
	\label{fig:Fig4}
\end{figure*}

As a demonstration of these techniques, we study Rydberg atoms that are coupled to the microwave electric field of the cavity in the presence of controlled dc electric and magnetic fields. Rydberg atoms, which are sent through the cavity via the access holes, are a sensitive probe for the applied electric and magnetic fields due to Stark and Zeeman effects, respectively. In the setup sketched in Fig.~\ref{fig:Fig4}(a) (see~\cite{Stammeier2017,Thiele2014} for details), in each experimental cycle, we prepare Rydberg atoms, apply a pulsed microwave tone to the cavity when the atoms are within the cavity and detect the atoms after they exited the cavity. 
A supersonic sample of ${{}^{4}\text{He}}^{*}$ atoms in the metastable triplet state ($1\text{s}^{1}2\text{s}^{1}~{}^{3}\text{S}_{1}$) travels at a speed of $v_{z}=900 \pm 13 ~\text{m}\thinspace{\text{s}}^{-1}$ in an ultrahigh vacuum, cryogenic environment at $3~\text{K}$, where the 3D cavity is mounted between two pairs of circular electrodes. Between the first pair of electrodes we employ two-photon excitation (see level diagram in~\ref{fig:Fig4}(b)) in a magnetic field of $B_{\text{exc}}\simeq 5~\text{G}$ using a $500$-$\text{ns}$-long laser pulse at wavelength $389~\text{nm}$ and a continuous-wave laser of wavelength $788~\text{nm}$ to transfer ${\text{He}}^{*}$ atoms to the $42\text{s}$ Rydberg state ($1\text{s}^{1}42\text{s}^{1}~{}^{3}\text{S}$) with spin and angular momentum projections $m_s = +1$ and $m_l = 0$, respectively. When the atom cloud arrives at the cavity center it has an approximate size of $\simeq 1.5~\text{mm}$ in both longitudinal and transversal  directions, determined by the laser pulse length, beam waist and the expansion after an 0.5-mm-diameter skimmer. Here, we coherently transfer a fraction of the  atoms to the $42\text{p}$ Rydberg states ($1\text{s}^{1}42\text{p}^{1}~{}^{3}\text{P}$) in the applied magnetic and electric fields by injecting a $500$-$\text{ns}$-long microwave pulse into the cavity. We use low microwave powers adjusted to compensate the frequency-dependent resonance curve of the cavity $\text{TE}_{301}$ mode, which is tuned to  $\nu_3 = 20.558~\text{GHz}$ by the insertion of a niobium cylinder. After leaving the cavity, the Rydberg atoms are ionized with a pulsed field at the second pair of electrodes and the resulting electrons are detected on a microchannel plate (MCP). The signal from the MCP, current-amplified and digitally integrated over a $70$-$\text{ns}$-long time window, creates a signal, $S$, which is state-dependent because of different field ionization thresholds for the s- and p-Rydberg states. It thus results in a spectral line $S(\nu)$, when the frequency $\nu$ of the microwave pulse is scanned over an atomic transition frequency.

The dc magnetic and electric fields applied in the cavity, $\boldsymbol{B}_{\text{dc}}$ and $\boldsymbol{E}_{\text{dc}}$, shift the atomic energy levels through the Zeeman effect (in the Paschen-Back regime) and the quadratic Stark effect, which are schematically outlined in Fig.~\ref{fig:Fig4}(c) and (d), respectively. 

The magnetic field is generated by a pair of 60-cm-diameter Helmholtz coils, aligned with the beam axis outside of the vacuum system of the experiment, and then trapped in the access holes during the cooldown of the cavity below the superconducting transition temperature $T_{\text{c}}\simeq 9.2~\text{K}$ of niobium. We can reset the trapped magnetic flux and thus the resulting field by heating the cavity to temperatures a few degrees above $T_{\text{c}}$, changing the current in the Helmholtz coils and cooling the cavity down again to the base temperature of $3~\text{K}$, an operation which takes about five minutes in our cryogenic system.

 In contrast, the dc electric field is aligned perpendicular to the atomic beam axis and can be changed  rapidly by applying potentials $V_1$ and $V_2$ to the intra-cavity electrodes. Indeed, electrostatic simulations show that the total capacitance of the  electrode $C\simeq 4~\text{pF}$ is dominated by the coaxial microwave cable (length $\simeq 30~\text{mm}$) needed for thermalization and connection, which leads to a rise time of $\simeq 0.4~\text{ns}$ (bandwidth $\simeq 800~\text{MHz}$), when driven with a voltage source of 50-$\Omega$ impedance.

In Fig.~\ref{fig:Fig4}(e) we study the dependence of atomic transition frequencies on the applied magnetic field. With compensated stray electric fields using potentials $V_{1} = 0~\text{V}$ and $V_{2} = 0.4~\text{V}$ at the intra-cavity electrodes we measure the normalized MCP signal $S$ (gray data points) as a function of the frequency detuning $\nu - \nu_0$ between the drive tone and the (magnetic and electric) field-free transition frequency $\nu_0 = 20.542~\text{GHz}$. In zero applied magnetic field we observe a single transition in the measured spectral line $S(\nu - \nu_0)$, which is Stark-shifted from $\nu_0$ by $\simeq -2.0~\text{MHz}$, because of a residual stray electric field of magnitude $E_{\text{res}}\simeq 100~\text{mV}/\text{cm}$.  
At magnetic fields above a threshold of $\simeq 3~\text{G}$ the data clearly displays two transitions and fits of the spectral lines with a double Gaussian (black lines) allow the extraction of the resonance frequencies, $\nu_{1}$ and $\nu_{-1}$ (blue and black data points), and full widths at half maximum (green bars) of the transitions to the $42\text{p}$ states with $m_l= 1$ and $m_l= -1$, respectively. The linear dependence of both transition frequencies in the applied magnetic field is in agreement with the Zeeman effect in the Paschen-Back regime, $\nu_{\pm 1} =\nu_{0\text{,}\pm 1} \pm \mu_{\text{B}} g_L B_{\text{dc}} /h$,  where $\mu_{\text{B}}$, $g_L \simeq 1$ and $h$ are the Bohr magneton, the orbital Land\'e g-factor and the Planck constant, respectively. 

Fitting the Zeeman effect to our data (red lines) we obtain a conversion relation of $5.1(1)~\text{G}/\text{A}$ between the magnetic field inside the cavity and the applied coil current. The relation agrees within the experimental uncertainty with the magnetic field per applied current analytically calculated for the used pair of Helmholtz coils. The fit also determines the zero-field offsets $\nu_{0\text{,}1}\simeq -2.5~\text{MHz}$ and $\nu_{0\text{,}-1}\simeq -0.5~\text{MHz}$, which have the same order of magnitude as the fine structure between the different J-states (maximal splitting $2.9~\text{MHz}$).  The width of the atomic spectral lines $\sigma_{\nu} \simeq 3.3(6)~\text{MHz}$ is determined by the inhomogeneous distribution of the residual electric field and shows no significant dependence on the magnetic field strength. This is in agreement with the magnetic field simulations shown in Fig.~\ref{fig:Fig1}(f), where the inhomogeneity of the magnetic field over the size of the atom cloud (estimated extension $\simeq 1.5~\text{mm}$) is less than one percent, which even for the highest applied magnetic field would lead to a negligible inhomogeneous broadening of $\sigma_{\nu \text{,}B}\simeq 0.1~\text{MHz}$.

In Fig.~\ref{fig:Fig4}(f) we present measurements of the transition frequencies, $\nu_{1}$ and $\nu_{-1}$ (blue and black data points), and the corresponding spectral linewidths (green bars) as a function of the applied dc electric field in a constant magnetic field of $B_{\text{dc}}\simeq 9.8~\text{G}$. Here, the offset potentials $V_{1,0} = 0.6~\text{V}$ and $V_{2,0} = 0.73~\text{V}$ at the intra-cavity electrodes minimize the residual stray electric field to $E_{\text{res}}\simeq 40~\text{mV}/\text{cm}$ as extracted from the Stark shift of $\simeq -0.4~\text{MHz}$. In this situation we increase only the potential of the second electrode by an amount $\Delta V_2$ ($V_{2}=V_{2,0} +\Delta V_2$), which leads to an applied electric field $E_{\text{dc}}=\beta \Delta V_{2}$ with a geometry-dependent factor $\beta$. We fit the shifts on both transition frequencies, $\Delta \nu_{\pm 1}(E_{\text{dc}}) = \nu_{\pm 1}(E_{\text{dc}}) - \nu_{\pm 1}(0)$ with the same quadratic Stark effect, $\Delta \nu(E_{\text{dc}}) = - 0.5 \delta_\alpha (\beta \Delta V_{2})^2$, (red lines), which allows the calibration of the electric fields indicated at the bottom axis using the calculated difference in polarizability $\delta_\alpha = -444~\text{MHz}/ ({\text{V}}/ {\text{cm}})^2$ between the 42s and 42p Rydberg states. The spatial distribution of the applied electric field causes a visible broadening $\sigma_{\nu \text{,}E} ={({\sigma_{\nu}(E_{\text{dc}})}^2 - {\sigma_{\nu}(0)}^2 )}^{1/2} $ of the atomic linewidth $\sigma_{\nu}(E_{\text{dc}})$, which exceeds the initial linewidth $\sigma_{\nu}(0)$ by a factor four for the highest electric fields. In this limit the inhomogeneous broadening is proportional to the Stark shift and allows the calculation of the relative field inhomogeneity $\sigma_{E} / E_{\text{dc}} = \sigma_{\nu \text{,}E} / (2\Delta \nu(E_{\text{dc}})) = 0.13(1)$. We then perform an electric field simulation (similar to the one shown in Fig.~\ref{fig:Fig1}(e)) using $V_{1} = 0~\text{V}$ and $V_{2} = 1~\text{V}$, which approximates the experimental conditions by neglecting the small stray electric field and the compensation thereof.
The experimental results for the proportionality factor $\beta = 0.67(1)~\text{cm}^{-1}$ between electric field and applied potential and the relative field inhomogeneity are both compatible with this simulation, when the simulated electric field distribution is averaged over an atomic cloud of transversal diameter $1.1~\text{mm}$ that is slightly smaller than previously estimated and offset from the cavity center by $0.7~\text{mm}$ in x-direction.

The presented measurements clearly show that dc electric and magnetic fields can be applied simultaneously in the center of the superconducting cavity facilitating the efficient control of the transition frequencies in Rydberg atoms, while maintaining a high cavity quality factor. The effect of dc electric fields on superconducting waveguide cavities was studied in~\cite{Cohen2017}, where electric fields of a few $\text{kV}/\text{cm}$ were shown to have no impact on the quality factor. A trapped magnetic field, $B_{\text{trap}}$, in the bulk material of the niobium cavity is expected to increase the surface resistivity by an amount $R_{\text{res}}=2.2~\text{n}\Omega / \mu \text{T}\cdot B_{\text{trap}} \sqrt{\nu/\text{GHz}}$ (see~\cite{Aull2012}) and thereby limit the quality factor. For the presented cavity we estimate a quality factor limited to  $10^6$ for a trapped magnetic field of $\simeq 20~\text{mT}$, which is an order of magnitude higher than the fields applied in our experiments. This makes the presented niobium cavity suitable for cavity QED experiments with superconducting qubits, NV centers~\cite{Angerer2016,Dolde2011} and other quantum emitters that can be controlled with magnetic fields below the critical magnetic field $B_{\text{c}}\simeq 200~\text{mT}$ of niobium.

\section{Cavity quality factor at the few photon level}

\begin{figure}[bt] \centering \includegraphics[width=85mm]{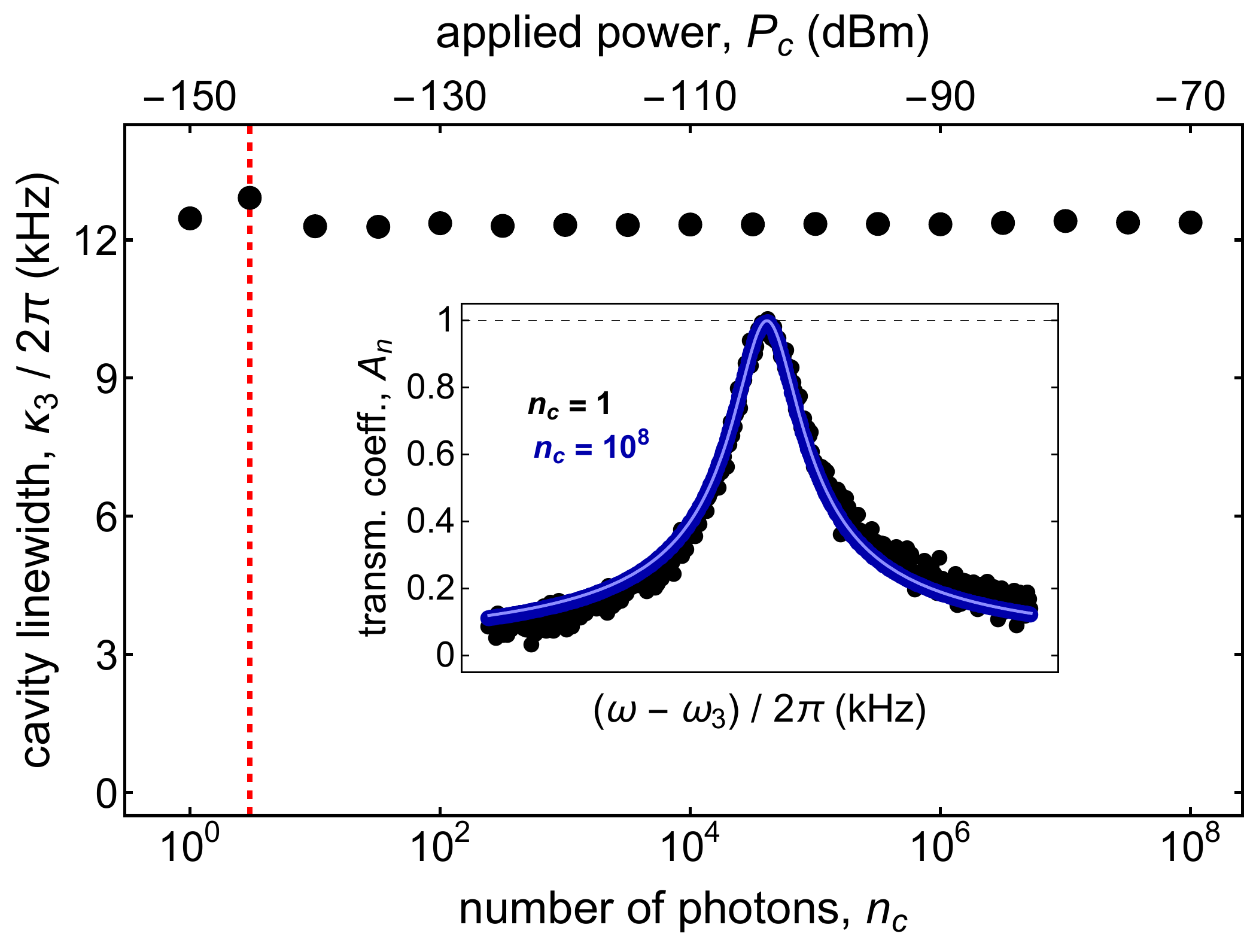} 	
	\caption{Photon number dependence of the cavity linewidth $\kappa_3 / 2\pi$ at $T = 3~\text{K}$ for the $\text{TE}_{301}$ mode. The photon number $n_{\text{c}}$ is calculated from input-output relations and the power $P_{\text{c}}$ applied to the cavity. The inset shows measurements of the normalized transmission coefficient $A_{\text{n}}$ as a function of the probe detuning $(\omega - \omega_3)/2\pi$ for the lowest ($n_{\text{c}}= 1 $, black) and the highest ($n_{\text{c}}= 10^8 $, blue) photon numbers populating the cavity. For better visibility only one Lorentzian fit (light blue) is shown. The red dashed line indicates the number of thermal photons in the cavity. 
	}
	\label{fig:Fig5}
\end{figure}

Cavity QED experiments typically require low photon numbers in the cavity, where cavity losses can be enhanced by the presence of unsaturated two-level systems~\cite{Bruno2015,Megrant2012}. To measure the photon number dependence of the internal losses in the $\text{TE}_{301}$ mode we mount the cavity, with 50-$\Omega$-terminated superconducting dc electrodes and the sapphire cylinder inserted by $1.6~\text{mm}$, into a cryogenic environment at $3~\text{K}$. We perform transmission measurements with thermalized attenuators and circulators on the input and output line, respectively, which provide sufficient thermalization of external blackbody radiation to have only about three thermal photons populating the cavity on average. The microwave signal transmitted through the cavity is amplified with a high-electron-mobility amplifier at $3~\text{K}$, before it undergoes further amplification and filtering at room temperature, followed by heterodyne downconversion using an analog mixer and digital homodyne downconversion (see~\cite{Stammeier2017} for details). In Fig.~\ref{fig:Fig5}, we show the dependence of the cavity linewidth $\kappa_3 /2\pi$ on the number of injected photons $n_{\text{c}}$, which is calculated from input-output relations and the power $P_{\text{c}}$ applied to the cavity. 
The cavity linewidths are determined from measurements of the normalized transmission coefficient $A_{\text{n}}$ as a function of the probe detuning $(\omega - \omega_3)/2\pi$, which are shown in the inset for the lowest ($n_{\text{c}}= 1 $, black) and the highest ($n_{\text{c}}= 10^8 $, blue) photon numbers. The data clearly shows a very narrow cavity linewidth of $\kappa_3 /2\pi = 12.4(1)~\text{kHz}$ (quality factor of $Q=1.65(1)\cdot 10^6$), which is independent of the photon number in the 3D cavity (in agreement with~\cite{Reagor2016}). The increase in quality factor compared to the previously shown measurements is due to improved thermalization of cavity and dc electrodes in the presented setup.

\FloatBarrier 

\section{Conclusions and Outlook}
 We conclude that the $\text{TE}_{301}$ mode of the superconducting cavity with holes for magnetic field access, electrodes for dc voltage bias and an inserted sapphire cylinder for resonance frequency tuning is able to maintain internal quality factors up to $Q_{\text{int}}\simeq 1.7\cdot 10^6$ at the few photon level, when it is thermalized at a temperature of $3~\text{K}$.

The measurements presented in this article show the feasibility of applying magnetic and electric field bias in superconducting cavities without compromising their quality factor at the few photon level. We believe that the presented results will be useful for cavity QED experiments with Rydberg atoms, superconducting qubits, NV-centers and opto-mechanical systems using 3D cavities. As one application, we have demonstrated the precise control of atomic transition frequencies in Rydberg atoms. We stress that the methods introduced here can most likely be adapted to apply magnetic and, more importantly, electric fields inside other three-dimensional cavity geometries, like for instance, cylindrical cavities. The simulated bandwidth ($\simeq 800~\text{MHz}$) for signals on the dc electrodes could also allow the manipulation of quantum systems with radio-frequency signals, such as e.g.~Rydberg Stark deceleration and trapping~\cite{Hogan2008} and ion trapping~\cite{Blatt2012}. Last but not least, we believe that the system can be adjusted to apply more complex electric and magnetic field configurations, since there is no obvious reason that prevents the insertion of multiple dc electrodes at the microwave electric field nodes and the use of multiple access holes in a superconducting cavity with quality factors of about one million.

\FloatBarrier 
\bigskip
\textbf{Acknowledgements}:

We thank Johannes Deiglmayr, Fr\'ed\'eric Merkt and Gianni Blatter for fruitful discussions.
We acknowledge the contributions of Tobias Thiele, Silvia Ruffieux, Andres Forrer and Bruno Eckmann to the initial phase of the experiment and thank the ETH physics machine shop. This work was supported by the European Union H$2020$ FET Proactive project RySQ (grant N. $640378$) and by the National Centre of Competence in Research "Quantum Science and Technology" (NCCR QSIT), a research instrument of the SNSF.

\appendix


\begin{thebibliography}{10}
	\providecommand{\url}[1]{\texttt{#1}}
	\providecommand{\urlprefix}{URL }
	\providecommand{\bibinfo}[2]{#2}
	\providecommand{\eprint}[2][]{\url{#2}}
	\providecommand \doibase [0]{http://dx.doi.org/}%
	

	



% QED
	\bibitem{Haroche2006}
	\bibinfo{author}{S.~Haroche, J.-M.~Raimond}, 
	\emph{\bibinfo{title}{Exploring the Quantum, 1st edition}}, 
	{\bibinfo{journal}{(Oxford University Press, Oxford, 2006)}}.	

% QIP

	\bibitem{Chuang2010}
	\bibinfo{author}{M.~A. Nielsen, I.~L. Chuang}, 
	\emph{\bibinfo{title}{Quantum Computation and Quantum Information, 10th edition}}, 
	{\bibinfo{journal}{(Cambridge University Press, Cambridge, 2010)}}.	


% cavity Qed experiments with Rydberg atoms	
	\bibitem{Haroche2013}
	\bibinfo{author}{S.~Haroche},
	%	\emph{\bibinfo{title}{Nobel Lecture: Controlling photons in a box and exploring the quantum to classical boundary*}},
	\href {\doibase 10.1103/RevModPhys.85.1083}
	{\bibinfo{journal}{Rev. Mod. Phys.} \textbf{\bibinfo{volume}{85}},
	\bibinfo{pages}{1083} (\bibinfo{year}{2013})}.


	\bibitem{Paik2011}
	\bibinfo{author}{H.~Paik}, \bibinfo{author}{D.~I. Schuster}, \bibinfo{author}{L.~S. Bishop}, \bibinfo{author}{G.~Kirchmair},
	\bibinfo{author}{G.~Catelani}, \bibinfo{author}{A.~P. Sears},	\bibinfo{author}{B.~R. Johnson},
	\bibinfo{author}{M.~J. Reagor}, \bibinfo{author}{L.~Frunzio}, \bibinfo{author}{L.~I. Glazman}, 
	\bibinfo{author}{S.~M. Girvin}, \bibinfo{author}{M.~H. Devoret}, and \bibinfo{author}{R.~J. Schoelkopf},
	% \emph{\bibinfo{title}{Observation of High Coherence in Josephson Junction Qubits Measured in a Three-Dimensional Circuit QED Architecture}}, 
	\href {\doibase 10.1103/PhysRevLett.107.240501} 
	{\bibinfo{journal}{Phys. Rev. Lett.} \textbf{\bibinfo{volume}{107}}, \bibinfo{pages}{240501 }
		(\bibinfo{year}{2011})}.
	
	
	\bibitem{Rigetti2012}
	\bibinfo{author}{C.~Rigetti}, \bibinfo{author}{J.~M. Gambetta},
	\bibinfo{author}{S.~Poletto}, \bibinfo{author}{B.~L.~T. Plourde},
	\bibinfo{author}{J.~M. Chow}, \bibinfo{author}{A.~D. Corcoles}, 
	\bibinfo{author}{J.~A. Smolin}, \bibinfo{author}{S.~T. Merkel},
	\bibinfo{author}{J.~R. Rozen}, \bibinfo{author}{G.~A. Keefe},
	\bibinfo{author}{M.~B. Rothwell}, \bibinfo{author}{M.~B. Ketchen},
	and \bibinfo{author}{M.~Steffen},
	%\emph{\bibinfo{title}{Superconducting qubit in a waveguide cavity with a coherence time approaching 0.1 ms}},
	\href {\doibase 10.1103/PhysRevB.86.100506}
	{\bibinfo{journal}{Phys. Rev. B} \textbf{\bibinfo{volume}{86}},
		\bibinfo{pages}{100506(R)} (\bibinfo{year}{2012})}.


% no photon number dependence at very high Q
	
	\bibitem{Reagor2013}
	\bibinfo{author}{M.~Reagor}, \bibinfo{author}{H.~Paik},
	\bibinfo{author}{G.~Catelani}, \bibinfo{author}{L.~Sun},
	\bibinfo{author}{C.~Axline}, \bibinfo{author}{E.~Holland}, \bibinfo{author}{I.~M. Pop},
	\bibinfo{author}{N.~A. Masluk}, \bibinfo{author}{T.~Brecht},
	\bibinfo{author}{L.~Frunzio}, \bibinfo{author}{M.~H. Devoret},
	\bibinfo{author}{L.~Glazman}, and \bibinfo{author}{R.~J. Schoelkopf},
	%	\emph{\bibinfo{title}{Reaching 10ms single photon lifetimes for superconducting aluminum cavities}}, 
	\href {\doibase 10.1063/1.4807015} 
	{\bibinfo{journal}{Appl. Phys. Lett.} \textbf{\bibinfo{volume}{102}}, \bibinfo{pages}{192604}
		(\bibinfo{year}{2013})}.
	

% Cavity QED with superconducting qubits (magnetic field important)

	\bibitem{Sun2014}
	\bibinfo{author}{L.~Sun}, \bibinfo{author}{A.~Petrenko}, \bibinfo{author}{Z.~Leghtas},
	\bibinfo{author}{B.~Vlastakis}, \bibinfo{author}{G.~Kirchmair}, 
	\bibinfo{author}{K.~M. Sliwa}, \bibinfo{author}{A.~Narla}, \bibinfo{author}{M.~Hatridge},
	\bibinfo{author}{S.~Shankar}, \bibinfo{author}{J.~Blumoff}, \bibinfo{author}{L.~Frunzio},
	\bibinfo{author}{M.~Mirrahimi},  \bibinfo{author}{M.~H. Devoret},
	and \bibinfo{author}{R.~J. Schoelkopf},
	%\emph{\bibinfo{title}{Tracking photon jumps with repeated quantum non-demolition parity measurements}}, 
	\href {\doibase 10.1038/nature13436}
	{\bibinfo{journal}{Nature (London)} \textbf{\bibinfo{volume}{511}},
		\bibinfo{pages}{444} (\bibinfo{year}{2014})}.


	\bibitem{Kirchmair2013}
	\bibinfo{author}{G.~Kirchmair}, \bibinfo{author}{B.~Vlastakis},
	\bibinfo{author}{Z.~Leghtas}, \bibinfo{author}{S.~E. Nigg},
	\bibinfo{author}{H.~Paik}, \bibinfo{author}{E.~Ginossar},
	\bibinfo{author}{M.~Mirrahimi}, \bibinfo{author}{L.~Frunzio}, \bibinfo{author}{S.~M. Girvin},
	and \bibinfo{author}{R.~J. Schoelkopf},
	%\emph{\bibinfo{title}{Observation of quantum state collapse and revival due to the single-photon Kerr effect}},
	 \href {\doibase 10.1038/nature11902}
	{\bibinfo{journal}{Nature (London)} \textbf{\bibinfo{volume}{495}},
	\bibinfo{pages}{205} (\bibinfo{year}{2013})}.


	\bibitem{Gasparinetti2016}
	\bibinfo{author}{S.~Gasparinetti}, \bibinfo{author}{S.~Berger},
	\bibinfo{author}{A.~A. Abdumalikov}, \bibinfo{author}{M.~Pechal},
	\bibinfo{author}{S.~Filipp}, and \bibinfo{author}{A.~Wallraff},
	%\emph{\bibinfo{title}{Measurement of a vacuum-induced geometric phase}},
	\href {\doibase 10.1126/sciadv.1501732}
	{\bibinfo{journal}{Sci. Adv.} \textbf{\bibinfo{volume}{2}},
	\bibinfo{pages}{e1501732} (\bibinfo{year}{2016})}.

	
	\bibitem{Flurin2015}
	\bibinfo{author}{E.~Flurin}, \bibinfo{author}{N.~Roch}, \bibinfo{author}{J.~D. Pillet}, \bibinfo{author}{F.~Mallet}, and \bibinfo{author}{B.~Huard},
	%\emph{\bibinfo{title}{Superconducting Quantum Node for Entanglement and Storage of Microwave Radiation}}, 
	\href {\doibase 10.1103/PhysRevLett.114.090503} 
	{\bibinfo{journal}{Phys. Rev. Lett.} \textbf{\bibinfo{volume}{114}}, \bibinfo{pages}{090503}
		(\bibinfo{year}{2015})}.	
	
	
	\bibitem{Heeres2015}
	\bibinfo{author}{R.~W. Heeres}, \bibinfo{author}{B.~Vlastakis}, \bibinfo{author}{E.~Holland}, \bibinfo{author}{S.~Krastanov},
	\bibinfo{author}{V.~V Albert}, \bibinfo{author}{L.~Frunzio}, \bibinfo{author}{L.~Jiang}, 
	and \bibinfo{author}{R.~J. Schoelkopf},
	%\emph{\bibinfo{title}{Cavity State Manipulation Using Photon-Number Selective Phase Gates}}, 
	\href {\doibase 10.1103/PhysRevLett.115.137002} 
	{\bibinfo{journal}{Phys. Rev. Lett.} \textbf{\bibinfo{volume}{115}}, \bibinfo{pages}{137002}
		(\bibinfo{year}{2015})}.	
	
	
	\bibitem{Reagor2016}
	\bibinfo{author}{M.~Reagor}, \bibinfo{author}{W.~Pfaff},
	\bibinfo{author}{C.~Axline}, \bibinfo{author}{R.~W. Heeres},
	\bibinfo{author}{N.~Ofek}, \bibinfo{author}{K.~Sliwa}, 
	\bibinfo{author}{E.~Holland}, \bibinfo{author}{C.~Wang},
	\bibinfo{author}{J.~Blumoff}, \bibinfo{author}{K.~Chou},
	\bibinfo{author}{M.~J. Hatridge}, \bibinfo{author}{L.~Frunzio},
	\bibinfo{author}{M.~H. Devoret}, \bibinfo{author}{L.~Jiang},
	and \bibinfo{author}{R.~J. Schoelkopf},
	%\emph{\bibinfo{title}{Quantum memory with millisecond coherence in circuit QED}},
	\href {\doibase 10.1103/PhysRevB.94.014506}
	{\bibinfo{journal}{Phys. Rev. B} \textbf{\bibinfo{volume}{94}},
		\bibinfo{pages}{014506} (\bibinfo{year}{2016})}.
	

	
	% Cavity QED with Spins (magnetic field important)
	
	\bibitem{Angerer2016}
	\bibinfo{author}{A.~Angerer}, \bibinfo{author}{T.~Astner}, \bibinfo{author}{D.~Wirtitsch}, \bibinfo{author}{H.~Sumiya},
	\bibinfo{author}{S.~Onoda}, \bibinfo{author}{J.~Isoya}, \bibinfo{author}{S.~Putz}, 
	and \bibinfo{author}{J.~Majer},
	%	\emph{\bibinfo{title}{Collective strong coupling with homogeneous Rabi frequencies using a 3D lumped element microwave resonator}}, 
	\href {\doibase 10.1063/1.4959095} 
	{\bibinfo{journal}{Appl. Phys. Lett.} \textbf{\bibinfo{volume}{109}}, \bibinfo{pages}{033508}
		(\bibinfo{year}{2016})}.
	
	
	\bibitem{Zhang2014}
	\bibinfo{author}{X.~Zhang}, \bibinfo{author}{C-L.~Zou}, \bibinfo{author}{L.~Jiang}, and \bibinfo{author}{H.~X. Tang},
	%	 \emph{\bibinfo{title}{Strongly Coupled Magnons and Cavity Microwave Photons}}, 
	\href {\doibase 10.1103/PhysRevLett.113.156401} 
	{\bibinfo{journal}{Phys. Rev. Lett.} \textbf{\bibinfo{volume}{113}}, \bibinfo{pages}{156401}
		(\bibinfo{year}{2014})}.
	
	
	\bibitem{Probst2014}
	\bibinfo{author}{S.~Probst}, \bibinfo{author}{A.~Tkalcec},
	\bibinfo{author}{H.~Rotzinger}, \bibinfo{author}{D.~Rieger},
	\bibinfo{author}{J-M.~Le~Floch}, \bibinfo{author}{M.~Goryachev}, 
	\bibinfo{author}{M.~E. Tobar}, \bibinfo{author}{A.~V. Ustinov},
	and \bibinfo{author}{P.~A. Bushev},
	%	\emph{\bibinfo{title}{Three-dimensional cavity quantum electrodynamics with a rare-earth spin ensemble}},
	\href {\doibase 10.1103/PhysRevB.90.100404}
	{\bibinfo{journal}{Phys. Rev. B} \textbf{\bibinfo{volume}{90}},
		\bibinfo{pages}{100404(R)} (\bibinfo{year}{2014})}.
	
	
	\bibitem{Tabuchi2015}
	\bibinfo{author}{Y.~Tabuchi}, \bibinfo{author}{S.~Ishino},
	\bibinfo{author}{A.~Noguchi},
	\bibinfo{author}{T.~Ishikawa}, \bibinfo{author}{R.~Yamazaki},
	\bibinfo{author}{K.~Usami}, and \bibinfo{author}{Y.~Nakamura},
	%	\emph{\bibinfo{title}{Coherent coupling between a ferromagnetic magnon and a superconducting qubit}},
	\href {\doibase 10.1126/science.aaa3693} {\bibinfo{journal}{Science} \textbf{\bibinfo{volume}{349}}, \bibinfo{pages}{405}
		(\bibinfo{year}{2015})}.
	
	

		
% electrically tunable graphene resonator
	
	\bibitem{Singh2014}
	\bibinfo{author}{V.~Singh}, \bibinfo{author}{S.~J. Bosman},
	\bibinfo{author}{B.~H. Schneider}, \bibinfo{author}{Y.~M. Blanter},
	\bibinfo{author}{A.~Castellanos-Gomez}, and \bibinfo{author}{G.~A. Steele},
	%\emph{\bibinfo{title}{Optomechanical coupling between a multilayer graphene mechanical resonator and a superconducting microwave cavity}},
	\href {\doibase 10.1038/NNANO.2014.168}
	{\bibinfo{journal}{Nature Nanotech.} \textbf{\bibinfo{volume}{9}},
		\bibinfo{pages}{820} (\bibinfo{year}{2014})}.
	
	
% Optomechanics with 3D cavity
	
	\bibitem{Yuan2015}
	\bibinfo{author}{M.~Yuan}, \bibinfo{author}{V.~Singh},
	\bibinfo{author}{Y.~M. Blanter}, and \bibinfo{author}{G.~A. Steele},
%	\emph{\bibinfo{title}{Large cooperativity and microkelvin cooling with a three-dimensional optomechanical cavity}}, 
	\href {\doibase 10.1038/ncomms9491}
	{\bibinfo{journal}{Nature Comm.} \textbf{\bibinfo{volume}{6}},
		\bibinfo{pages}{8491} (\bibinfo{year}{2015})}.
	
	
% electric field dependence of NV centers	

	\bibitem{Dolde2011}
	\bibinfo{author}{F.~Dolde}, \bibinfo{author}{H.~Fedder},
	\bibinfo{author}{M.~W. Doherty}, \bibinfo{author}{T.~N\"obauer},
	\bibinfo{author}{F.~Rempp}, \bibinfo{author}{G.~Balasubramanian}, \bibinfo{author}{T.~Wolf},
		\bibinfo{author}{F.~Reinhard}, 	\bibinfo{author}{L.~C.~L. Hollenberg}, 	\bibinfo{author}{F.~Jelezko},
		and \bibinfo{author}{J.~Wrachtrup},
	%\emph{\bibinfo{title}{Electric-field sensing using single diamond spins}},
	\href {\doibase 10.1038/NPHYS1969}
	{\bibinfo{journal}{Nature Phys.} \textbf{\bibinfo{volume}{7}},
		\bibinfo{pages}{459} (\bibinfo{year}{2011})}.
	
	
	\bibitem{Hare1988}
	\bibinfo{author}{J.~Hare}, \bibinfo{author}{M.~Gross}, and \bibinfo{author}{P.~Goy},
	% \emph{\bibinfo{title}{Circular Atoms Prepared by a New Method of Crossed Electric and Magnetic Fields}}, 
	\href {\doibase 10.1103/PhysRevLett.61.1938} 
	{\bibinfo{journal}{Phys. Rev. Lett.} \textbf{\bibinfo{volume}{61}}, \bibinfo{pages}{1938}
		(\bibinfo{year}{1988})}.
	
	
% Bfield in 3D cavity	

	\bibitem{Reshitnyk2016}
	\bibinfo{author}{Y.~Reshitnyk}, \bibinfo{author}{M.~Jerger}, and \bibinfo{author}{A.~Fedorov},
	%	\emph{\bibinfo{title}{3D microwave cavity with magnetic flux control and enhanced quality factor}}, 
	\href {\doibase 10.1140/epjqt/s40507-016-0050-8}
	{\bibinfo{journal}{EPJ Quantum Technol.} \textbf{\bibinfo{volume}{3}},
		\bibinfo{pages}{13} (\bibinfo{year}{2016})}.
	

	
% Efield in 3D cavity
	
	\bibitem{Kong2015}
	\bibinfo{author}{W-C.~Kong}, \bibinfo{author}{G-W.~Deng},
	\bibinfo{author}{S-X.~Li}, \bibinfo{author}{H-O.~Li},
	\bibinfo{author}{G.~Cao}, \bibinfo{author}{M.~Xiao}, and \bibinfo{author}{G-P.~Guo},
	%	\emph{\bibinfo{title}{Introduction of DC line structures into a superconducting microwave 3D cavity}}, 
	\href {\doibase 10.1063/1.4913252}
	{\bibinfo{journal}{Rev. Sci. Instrum.} \textbf{\bibinfo{volume}{86}},
		\bibinfo{pages}{023108} (\bibinfo{year}{2015})}.	
	
	
	\bibitem{Cohen2017}
	\bibinfo{author}{M.~A. Cohen}, \bibinfo{author}{M.~Yuan},
	\bibinfo{author}{B.~W.~A. de Jong}, \bibinfo{author}{E.~Boukers},
	\bibinfo{author}{S.~J. Bosman}, and \bibinfo{author}{G.~A. Steele},
	%	\emph{\bibinfo{title}{A split-cavity design for the incorporation of a DC bias in a 3D microwave cavity}}, 
	\href {\doibase 10.1063/1.4981884} 
	{\bibinfo{journal}{Appl. Phys. Lett.} \textbf{\bibinfo{volume}{110}}, \bibinfo{pages}{172601}
		(\bibinfo{year}{2017})}.
	

	\bibitem{HFSS2014}
	\bibinfo{author}{ANSYS\textregistered~HFSS\texttrademark~2014}
%	{\bibinfo{title}{High frequency structure solver (HFSS) 2014}}
	\href {http://www.ansys.com/products/electronics/rf-and-microwave} .

	\bibitem{Maxwell2014}
	\bibinfo{author}{ANSYS\textregistered~Maxwell\texttrademark~2014}
%	{\bibinfo{title}{High frequency structure solver (HFSS) 2014}}
	\href {http://www.ansys.com/products/electronics/rf-and-microwave} .
	
	
	% reference for trapped magnetic flux in niobium	
	
	\bibitem{Aull2012}
	\bibinfo{author}{S.~Aull}, \bibinfo{author}{O.~Kugeler}, and \bibinfo{author}{J.~Knobloch},
	%	\emph{\bibinfo{title}{Trapped magnetic flux in superconducting niobium samples}}, 
	\href {\doibase 10.1103/PhysRevSTAB.15.062001} 
	{\bibinfo{journal}{Phys. Rev. STAB} \textbf{\bibinfo{volume}{15}}, \bibinfo{pages}{062001}
		(\bibinfo{year}{2012})}.
	
	
	\bibitem{Pozar2011}
	\bibinfo{author}{D.~M. Pozar}, 
	\emph{\bibinfo{title}{Microwave Engineering, 4th edition}}, 
	{\bibinfo{journal}{(John Wiley and Sons Inc., New York, 2011)}}.	
	
	\bibitem{Timmerhaus1977}
	\bibinfo{author}{K.~D. Timmerhaus, R.~P. Reed and A.~F. Clark}, 
	\emph{\bibinfo{title}{Adavances in Cryogenic Engineering, Volume 22}}, 
	{\bibinfo{journal}{(Plenum Press, New York, 1977)}}.	
	
	% Tuning of superconducting cavity
	
	\bibitem{Souris2017}
	\bibinfo{author}{F.~Souris}, \bibinfo{author}{H.~Christiani},
	and \bibinfo{author}{J.~P. Davis},
	%	\emph{\bibinfo{title}{Tuning a 3D microwave cavity via superfluid helium at millikelvin temperatures}}, 
	\href {\doibase 10.1063/1.4997641} 
	{\bibinfo{journal}{Appl. Phys. Lett.} \textbf{\bibinfo{volume}{111}}, \bibinfo{pages}{172601}
	(\bibinfo{year}{2017})}.	
	
	
	\bibitem{Carvalho2016}
	\bibinfo{author}{N.~C. Carvalho}, \bibinfo{author}{Y.~Fan},
	and \bibinfo{author}{M.~E. Tobar},
	%	\emph{\bibinfo{title}{Piezoelectric tunable microwave superconducting cavity}}, 
	\href {\doibase 10.1063/1.4962695}
	{\bibinfo{journal}{Rev. Sci. Instrum.} \textbf{\bibinfo{volume}{87}},
		\bibinfo{pages}{094702} (\bibinfo{year}{2016})}.
	


	
	
% 3D cavity as atom number detector
	
	\bibitem{Stammeier2017}
	\bibinfo{author}{M.~Stammeier}, \bibinfo{author}{S.~Garcia},
	\bibinfo{author}{T.~Thiele}, \bibinfo{author}{J.~Deiglmayr},
	\bibinfo{author}{J.~A. Agner}, \bibinfo{author}{H.~Schmutz}, 
	\bibinfo{author}{F.~Merkt}, and \bibinfo{author}{A.~Wallraff},
%	\emph{\bibinfo{title}{Measuring the dispersive frequency shift of a rectangular microwave cavity induced by an ensemble of Rydberg atoms}},
	\href {\doibase 10.1103/PhysRevA.95.053855}
	{\bibinfo{journal}{Phys. Rev. A} \textbf{\bibinfo{volume}{95}},
		\bibinfo{pages}{053855} (\bibinfo{year}{2017})}.
	
% setup reference could also use one more reference for Stark effect, electric fields
	
	\bibitem{Thiele2014}
	\bibinfo{author}{T.~Thiele}, \bibinfo{author}{S.~Filipp},
	\bibinfo{author}{J.~A. Agner}, \bibinfo{author}{H.~Schmutz},
	\bibinfo{author}{J.~Deiglmayr}, \bibinfo{author}{M.~Stammeier},
	\bibinfo{author}{P.~Allmendinger}, \bibinfo{author}{F.~Merkt},
	and \bibinfo{author}{A.~Wallraff},
%	\emph{\bibinfo{title}{Manipulating Rydberg atoms close to surfaces at cryogenic temperatures}}, 
	\href {\doibase 10.1103/PhysRevA.90.013414}
	{\bibinfo{journal}{Phys. Rev. A} \textbf{\bibinfo{volume}{90}},
	\bibinfo{pages}{013414} (\bibinfo{year}{2014})}.	
	



% reference for photon dependent Q-factor in CPW resonators	

	\bibitem{Bruno2015}
	\bibinfo{author}{A.~Bruno}, \bibinfo{author}{G.~de Lange},
	\bibinfo{author}{S.~Assad},	\bibinfo{author}{K.~L. van der Enden},
	 \bibinfo{author}{N.~K. Langford},  and \bibinfo{author}{L.~DiCarlo},
	%	\emph{\bibinfo{title}{Reducing intrinsic loss in superconducting resonators by surface treatment and deepetching of silicon substrates}}, 
	\href {\doibase 10.1063/1.4919761} 
	{\bibinfo{journal}{Appl. Phys. Lett.} \textbf{\bibinfo{volume}{106}}, \bibinfo{pages}{182601}
		(\bibinfo{year}{2015})}.


	\bibitem{Megrant2012}
	\bibinfo{author}{A.~Megrant}, \bibinfo{author}{C.~Neill},
	\bibinfo{author}{R.~Barends},	\bibinfo{author}{B.~Chiaro},
	\bibinfo{author}{Y.~Chen}, \bibinfo{author}{L.~Feigl},
	\bibinfo{author}{J.~Kelly},		\bibinfo{author}{E.~Lucero},	
	\bibinfo{author}{M.~Mariantoni},  \bibinfo{author}{P.~J.~J. O'Malley},
	\bibinfo{author}{D.~Sank},	\bibinfo{author}{A.~Vainsencher},		
	\bibinfo{author}{J.~Wenner}, 	\bibinfo{author}{T.~C. White},	
	\bibinfo{author}{Y.~Yin},		\bibinfo{author}{J.~Zhao},		
	\bibinfo{author}{C.~J. Palmstr\o m}, 	\bibinfo{author}{J.~M. Martinis},	 and \bibinfo{author}{A.~N. Cleland},
	%	\emph{\bibinfo{title}{Reducing intrinsic loss in superconducting resonators by surface treatment and deepetching of silicon substrates}}, 
	\href {\doibase 10.1063/1.3693409} 
	{\bibinfo{journal}{Appl. Phys. Lett.} \textbf{\bibinfo{volume}{100}}, \bibinfo{pages}{113510 }
	(\bibinfo{year}{2012})}.



%Rydberg Stark trapping
	
\bibitem{Hogan2008}
\bibinfo{author}{S.~D. Hogan} and \bibinfo{author}{F.~Merkt},
%\emph{\bibinfo{title}{Demonstration of three-dimensional electrostatic trapping of state-selected Rydberg atoms}}, 
\href {\doibase 10.1103/PhysRevLett.100.043001} 
{\bibinfo{journal}{Phys. Rev. Lett.} \textbf{\bibinfo{volume}{100}}, \bibinfo{pages}{043001}
	(\bibinfo{year}{2008})}.	

	
\bibitem{Blatt2012}
\bibinfo{author}{R.~Blatt}, and \bibinfo{author}{C.~F. Roos},
%\emph{\bibinfo{title}{Quantum simulations with trapped ions}},
\href {\doibase 10.1038/nphys2252}
{\bibinfo{journal}{Nature Phys.} \textbf{\bibinfo{volume}{8}},
	\bibinfo{pages}{277} (\bibinfo{year}{2012})}.	

% not used references	
	
% Efield in CPW resonator

%	\bibitem{Chen2011}
%	\bibinfo{author}{F.~Chen}, \bibinfo{author}{A.~J. Sirois},
%	\bibinfo{author}{R.~W. Simmonds}, and \bibinfo{author}{A.~J. Rimberg},
%	%	\emph{\bibinfo{title}{Introduction of a dc bias into a high-Q superconducting microwave cavity}}, 
%	\href {\doibase 10.1063/1.3573824} 
%	{\bibinfo{journal}{Appl. Phys. Lett.} \textbf{\bibinfo{volume}{98}}, \bibinfo{pages}{132509}
%		(\bibinfo{year}{2011})}.
	
	
	
%\bibitem{Novikov2013}
%\bibinfo{author}{S.~Novikov}, \bibinfo{author}{J.~E. Robinson},
%\bibinfo{author}{Z.~K. Keane}, \bibinfo{author}{B.~Suri},
%\bibinfo{author}{F.~C. Wellstood}, and \bibinfo{author}{B.~S. Palmer},
%\emph{\bibinfo{title}{Autler-Townes splitting in a three-dimensional transmon superconducting qubit}},
% \href {\doibase 10.1103/PhysRevB.88.060503}
%{\bibinfo{journal}{Phys. Rev. B} \textbf{\bibinfo{volume}{88}},
%	\bibinfo{pages}{060503(R)} (\bibinfo{year}{2013})}.

%\bibitem{Riste2013}
%\bibinfo{author}{D.~Riste}, \bibinfo{author}{M.~Dukalski},
%\bibinfo{author}{C.~A. Watson}, \bibinfo{author}{G.~de Lange},
%\bibinfo{author}{M.~J. Tiggelman}, \bibinfo{author}{Ya.~M. Blanter},
%\bibinfo{author}{K.~W. Lehnert}, \bibinfo{author}{R.~N. Schouten}, 
%and \bibinfo{author}{L.~DiCarlo},
%	%\emph{\bibinfo{title}{Deterministic entanglement of superconducting qubits by parity measurement and feedback}}, 
%	\href {\doibase 10.1038/nature12513}
%{\bibinfo{journal}{Nature (London)} \textbf{\bibinfo{volume}{502}},
%	\bibinfo{pages}{350} (\bibinfo{year}{2013})}.


%	\bibitem{Martinez2016}
%	\bibinfo{author}{L.~A. Martinez}, \bibinfo{author}{A.~R. Castelli},
%	\bibinfo{author}{W.~Delmas}, \bibinfo{author}{J.~E. Sharping}, and \bibinfo{author}{R.~Chiao},
%	\emph{\bibinfo{title}{Electromagnetic coupling to centimeter-scale mechanical membrane resonators via RF cylindrical cavities}}, 
%	\href {\doibase 10.1088/1367-2630/18/11/113015}
%	{\bibinfo{journal}{New J. Phys.} \textbf{\bibinfo{volume}{18}},
%	\bibinfo{pages}{113015} (\bibinfo{year}{2016})}.	
	
		
%	\bibitem{deGraaf2014}
%	\bibinfo{author}{S.~E. de Graaf}, \bibinfo{author}{D.~Davidovikj},
%	\bibinfo{author}{A.~Adamyan}, \bibinfo{author}{S.~E. Kubatkin}, and \bibinfo{author}{A.~V. Danilov},
%	\emph{\bibinfo{title}{Galvanically split superconducting microwave resonators for introducing internal voltage bias}}, 
%	\href {\doibase 10.1063/1.4863681} 
%	{\bibinfo{journal}{Appl. Phys. Lett.} \textbf{\bibinfo{volume}{104}}, \bibinfo{pages}{052601}
%		(\bibinfo{year}{2014})}.
	
%	\bibitem{Bosman2015}
%	\bibinfo{author}{S.~J. Bosman}, \bibinfo{author}{V.~Singh},
%	\bibinfo{author}{A.~Bruno}, and \bibinfo{author}{G.~A. Steele},
%	\emph{\bibinfo{title}{Broadband architecture for galvanically accessible superconducting microwave resonators}}, 
%	\href {\doibase 10.1063/1.4935346} 
%	{\bibinfo{journal}{Appl. Phys. Lett.} \textbf{\bibinfo{volume}{107}}, \bibinfo{pages}{192602}
%		(\bibinfo{year}{2015})}.
	


	
	
	
	% Bfield tuning of Transmon outside of 3D cavity
	
%	\bibitem{Juliusson2016}
%	\bibinfo{author}{K.~Juliusson}, \bibinfo{author}{S.~Bernon},
%	\bibinfo{author}{X.~Zhou}, \bibinfo{author}{V.~Schmitt},
%	\bibinfo{author}{H.~le Sueur}, \bibinfo{author}{P.~Bertet},
%	\bibinfo{author}{D.~Vion}, \bibinfo{author}{M.~Mirrahimi}, \bibinfo{author}{P.~Rouchon},
%	and \bibinfo{author}{D.~Esteve},
%	\emph{\bibinfo{title}{Manipulating Fock states of a harmonic oscillator while preserving its linearity}}, 
%	\href {\doibase 10.1103/PhysRevA.94.063861}
%	{\bibinfo{journal}{Phys. Rev. A} \textbf{\bibinfo{volume}{94}},
%		\bibinfo{pages}{063861} (\bibinfo{year}{2016})}.
	
	
	% Stark and Zeeman effect
	
%	\bibitem{Zimmerman1979}
%	\bibinfo{author}{M.~L. Zimmerman}, \bibinfo{author}{M.~G. Littman},
%	\bibinfo{author}{M.~M. Kash}, and \bibinfo{author}{D.~Kleppner},
%	\emph{\bibinfo{title}{Stark structure of the Rydberg states of alkali-metal atoms}}, 
%	\href {\doibase 10.1103/PhysRevA.20.2251}
%	{\bibinfo{journal}{Phys. Rev. A} \textbf{\bibinfo{volume}{20}},
%		\bibinfo{pages}{2251} (\bibinfo{year}{1979})}.	
		
%	\bibitem{Foot}
%	\bibinfo{author}{C.~J. Foot}, 
%	\emph{\bibinfo{title}{Atomic Physics}}, 
%	{\bibinfo{journal}{(Oxford University press, Oxford, 2005)}}.	
	
	
\end{thebibliography}
\end{document}